\documentclass[review]{elsarticle}

\usepackage{lineno,hyperref,amsmath}
\modulolinenumbers[5]

\journal{Templates of Journal }

\begin{document}

\begin{frontmatter}

\title{ {Improved GM(1,1) model based on} Simpson formula and its applications
 \tnoteref{mytitlenote}}

\tnotetext[mytitlenote]{To cite this document: Ma Xin, Wu Wenqing,
Zhang Yuanyuan, Improved GM(1,1) model based on Simpson formula and
its applications, Journal of Grey System, 2019, 31(4): 33-46. }

 \author[label1]{Xin Ma \fnref{label2}}
 \address[label1]{School of Science, Southwest University of Science and Technology, Mianyang 621010,P.R.China}
 \address[label2]{State Key Laboratory of Oil and Gas Reservoir Geology and Exploitation, Southwest Petroleum University, Chengdu 610500, P.R. China}
 \author[label1]{Wenqing Wu \fnref{label3} \corref{cor1}}
 \cortext[cor1]{Corresponding author. Wenqing Wu}
 \address[label3]{V.C. \& V.R. Key Lab of Sichuan Province, Sichuan Normal University, Chengdu 610068, P.R. China}
 \ead{wwqing0704@163.com}
 \author[label1]{Yuanyuan Zhang }
\begin{abstract}
The classical GM(1,1) model is an efficient tool to {make accurate
forecasts} with limited samples. But the accuracy of the GM(1,1)
model still needs to be improved. This paper proposes a novel
discrete GM(1,1) model, named ${\rm GM_{SD}}$(1,1) model, {of which
the background value is reconstructed using Simpson formula. The
expression of the specific time response function is deduced, and
the relationship between our model} and the continuous GM(1,1) model
with Simpson formula called ${\rm GM_{SC} }$(1,1) model {is
systematically discussed. The proposed model is proved to be}
unbiased to simulate the homogeneous exponent sequence. {Further,
some numerical examples are given to validate the accuracy of the
new ${\rm GM_{SD}}$(1,1) model. Finally, this model is used to
predict the Gross Domestic Product and the freightage of Lanzhou,
and the results illustrate the ${\rm GM_{SD}}$(1,1) model provides
accurate prediction.}
\end{abstract}
\end{frontmatter}

\section{Introduction}
\label{sec:introduction}

Grey system theory is an effective method to analyze uncertain
problems with small samples and poor information, which was founded
by professor Deng Julong \cite{Deng1982ControlSCL}. The principle of
the grey theory is ``grey box" of which some information is known
and the rest is unknown. Comparing with other methods, {such as
knowledge-driven method (Xiao et al. \cite{Xiao2019knowledgeE}),
fuzzy systems (Wu et al. \cite{Wu2017TopologicalIJBC}), hybrid
forecasting system (Du et al. \cite{Du2018multistepRE}, Ma et al.
\cite{Ma2019ChinaSTTE}), coupling mathematical model (Wang et al.
\cite{Wang2017TransientJPM, Wang2018FlowIJNSNS}),} the grey model
only needs little origin data having simple calculation process and
satisfactory forecasting accuracy. Due to this important feature, it
has been successfully applied in various fields. Its applications
include, but are not limited to, the inverted pendulum control
(Huang and Huang \cite{Huang2000ControlIEEETOIA}), the semiconductor
manufacturing layout (Chang et al. \cite{Chang2005ATFSC}), the stock
price forecasting (Chen et al. \cite{Chen2010ForecastingESA}), {the
energy production (Wang et al. \cite{Wang2017GreyAMM,
Wang2017DecompositionEP}, Zeng et al. \cite{Zeng2018ForecastingE},
Zhou and He \cite{Zhou2013GeneralizedAMM}), the energy consumption
(Ma and Liu \cite{Ma2017TheJGS}, Wu et al. \cite{Wu2018ApplicationE,
Wu2019ForecastingRE}), the industrial pollutant emission (Ma et al.
\cite{Ma2019anovelE}), the health expenditure of China (Wu et al.
\cite{Wu2019anallysisGSTA}), and China's electricity consumption
(Zeng \cite{Zeng2018ForecastingJGS}, Wu et al.
\cite{Wu2018usingE}).}

In 1982, Deng presented the classical continuous GM(1,1) model of
which procedures start with a differential equation called whitening
equation. By discretizing the whitening equation and employing the
least squares method, system parameters are estimated. Then
simulation values and prediction values are computed with the help
of the {whitening} equation and system parameters. Over the past
three decades, a great number of univariate grey forecasting models
have been proposed based on Deng's pioneer work. Some excellent
models in this area are those NNGBM(1,1) (Chen et al.
\cite{Chen2010ForecastingESA}, Zhang et al. \cite{Zhang2014AnCBM}),
GGM(1,1) (Zhou and He \cite{Zhou2013GeneralizedAMM}), DGM(1,1) (Xie
and Liu \cite{Xie2009DiscreteAMM}, Zeng et al.
\cite{Zeng2019AnewAPM}), NGBM(1,1) (Chen et al.
\cite{Chen2008ForecastingCNSNS}, Kong and Ma
\cite{kong2018comparisonGSTA}, Pei et al. \cite{pei2018theGSTA},
Salehi and Dehnavi \cite{Salehi2018AuditGSTA} ), SAGM(1,1) (Truong
and Ahn \cite{Truong2012AnESWA}), NGM(1,1,$k$) (Cui et al.
\cite{Cui2013AAPM}, Zeng and Li \cite{Zeng2018improvedCIE}).
Recently, He and Wang \cite{He2011ConstructingJQE} studied the
continuous GM(1,1) model, i.e., ${\rm GM_{SC}}$(1,1) model where the
background value was derived by utilizing the Simpson numerical
integration formula. But their model has been shown inaccurate in
some applications and biased for the homogeneous exponent sequence.
Thus the optimization of the grey model and the improvement of the
grey system theory have acquired a lot of attention. For instance,
an efficient way to improve the effectiveness of the grey models was
thw development of discrete grey models DGM(1,1) (Xie and Liu
\cite{Xie2009DiscreteAMM}). For more details, the readers are
directed to Xia et al. \cite{Xia2020ApplicationJCP}, Ma
\cite{Ma2019AJGS} and Wang and Phan \cite{Wang2015AnMPIE}.

In this paper, we focus on a kind of discrete GM(1,1) model called
${\rm GM_{SD}}$(1,1) model where the background value is computed
employing the Simpson numerical integration formula. Its solutions
of {time response function} and restored values, properties, and
applications are derived. We also study the forecast stability
problem of the discrete ${\rm GM_{SD}}$(1,1) model and discuss its
causes from continuous to discrete in detail. That our model is also
unbiased to simulate the homogeneous exponent sequence is proved.
Finally, we simulate and forecast the Gross Domestic Product and the
freightage of {Lanzhou} by using four kinds of GM(1,1) models.

The remainder of this paper is organized as follows. Section
\ref{sec:the-basis} gives a brief overview of the continuous GM(1,1)
model. Its solutions and properties of ${\rm GM_{SD}}$(1,1) model
are derived in Section \ref{sec:discreteGMSD}. {Section
\ref{sec:validation} discusses the validation of the ${\rm
GM_{SD}}$(1,1) model. Applications are provided in Section
\ref{sec:applications-example}.} Conclusions are drawn in Section 6.
\section{The basis of GM(1,1) model}
\label{sec:the-basis}

This section gives a brief overview of the classical continuous
GM(1,1) model. Suppose an original non-negative series be $X^{(0)}\!
= \! \left( {x^{(0)} (1),x^{(0)} (2), \ldots ,x^{(0)} (n)} \right)$
and the $x^{(0)} (k)$ represents the behavior of the data at the
time index $k$ for $k=1,2,\ldots,n$. Deng \cite{Deng1982ControlSCL}
proposed the GM(1,1) model is the following linear differential
equation
\begin{eqnarray}
\frac{{d x^{(1)} (t)}}{{dt}} + a x^{(1)} (t) = b,
\label{Eq:GM(1,1)-deng}
\end{eqnarray}
where the $x^{(1)} (k) = \sum\limits_{i = 1}^k {x^{(0)} (i)} ,\ k =
1,2, \ldots ,n$ are the {first-order} accumulated generating
operating (1-AGO) series of $X^{(0)}$, the $a$ and $b$ are system
parameters. Eq. (\ref{Eq:GM(1,1)-deng}) is also called the whitening
equation of the GM(1,1) model.

The approximation of $\frac{{dx^{(1)} (t)}}{{dt}}$ is taken as
\begin{eqnarray}
\frac{{dx^{(1)} (t)}}{{dt}} = \mathop {\lim }\limits_{\Delta t \to
1} \frac{{x^{(1)} (t) - x^{(1)} (t - \Delta t)}}{{\Delta t}} \approx
x^{(1)} (t) - x^{(1)} (t - 1) = x^{(0)} (t), \nonumber
\end{eqnarray}
and the background values of $x^{(1)} (t)$ are defined as
\begin{eqnarray}
z^{(1)} (t) = \frac{1}{2}\left( {x^{(1)} (t) + x^{(1)} (t - 1)}
\nonumber \right).
\end{eqnarray}

Thus the differential Eq.(\ref{Eq:GM(1,1)-deng}) can be
approximately rewritten as the following difference equation
\begin{eqnarray} \label{Eq:approximately-from-GM(1,1)}
x^{(0)} (t) + a z^{(1)} (t) = b.
\end{eqnarray}

Employing the least squares estimation method, from
Eq.(\ref{Eq:approximately-from-GM(1,1)}) by considering
$t=2,3,\ldots,n$, the model parameters $a$ and $b$ can be given
below
\begin{eqnarray}
\left( {\begin{array}{*{20}c}
   {\hat a}  \\
   {\hat b}  \\
\end{array}} \right)
 = \left( {\boldsymbol\Lambda^\mathcal{T} \boldsymbol\Lambda} \right)^{ - 1} \boldsymbol\Lambda^\mathcal{T}
 \boldsymbol\eta, \label{Eq:least-squares-GM(1,1)}
\end{eqnarray}
where $\boldsymbol\Lambda$ and $\boldsymbol\eta$ are defined as
follow
\begin{eqnarray}
\boldsymbol\Lambda = \left( {\begin{array}{*{20}cc}
   { - z^{(1)} (2)} && 1  \\
   { - z^{(1)} (3)} && 1  \\
        \vdots      &&  \vdots   \\
   { - z^{(1)} (\nu)} && 1  \\
\end{array}} \right), \quad
\boldsymbol\eta  = \left( {\begin{array}{*{20}c}
   {x^{(0)} (2)}  \\
   {x^{(0)} (3)}  \\
    \vdots   \\
   {x^{(0)} (\nu)}  \\
\end{array}} \right), \nonumber
\end{eqnarray}
where $\nu$ is the number of samples that are used to build the grey
models, and the left $n-\nu$ samples are used to test.

Solving Eq.(\ref{Eq:GM(1,1)-deng}), the time response {function can
be expressed by}
\begin{eqnarray}\label{Eq:response-sequence-GM(1,1)}
\hat x^{(1)} (k + 1) = \left( {x^{(0)} (1) - \frac{b}{a}} \right)e^{
- ak}  + \frac{b}{a},\ k=1,2,\ldots,n-1.
\end{eqnarray}

Then the restored values of $\hat x^{(0)} \left( {k + 1} \right)$
can be estimated by inverse accumulated generating operation (IAGO)
which is given by
\begin{eqnarray}\label{Eq:restored-values-GM(1,1)-a}
\hat x^{(0)} \left( {k + 1} \right) = \hat x^{(1)} \left( {k + 1}
\right) - \hat x^{(1)} \left( k \right),\ k=1,2,\ldots,n-1,
\end{eqnarray}
or
\begin{eqnarray}\label{Eq:restored-values-GM(1,1)-b}
\hat x^{(0)} (k + 1) = \frac{{e^a  - 1}}{a}\left( {b - a x^{(0)}
(1)} \right)e^{ - ak},\ k=1,2,\ldots,n-1.
\end{eqnarray}

As presented above, once given the sample data, the system
parameters in Eq.(\ref{Eq:GM(1,1)-deng}) obtained. The output series
also predicted with system parameters and input series by the
Eqs.(\ref{Eq:response-sequence-GM(1,1)})-(\ref{Eq:restored-values-GM(1,1)-b}).
\section{The discrete ${\rm GM_{SD}}$(1,1) model}
\label{sec:discreteGMSD}

\subsection{Representation of the discrete ${\rm GM_{SD}}$(1,1) model}
\label{subsec:representation}

This subsection derives the discrete ${\rm GM_{SD}}$(1,1) model with
Eq.(\ref{Eq:GM(1,1)-deng}) and the Simpson numerical integration
formula. Considering the integration of Eq.(\ref{Eq:GM(1,1)-deng})
in the interval $[k-1,\ k+1]$, it follows
\begin{eqnarray}\label{Eq:integration-[k-1, k+1]}
\int_{k - 1}^{k + 1} d x^{(1)} (t) + a\int_{k - 1}^{k + 1} {x^{(1)}
(t)} dt = b\int_{k - 1}^{k + 1} {dt},\ k=2,3,\ldots,n-1.
\end{eqnarray}

It follows from Eq.(\ref{Eq:integration-[k-1, k+1]}) that
\begin{eqnarray}
x^{(1)} (k + 1) - x^{(1)} (k - 1) + a\int_{k - 1}^{k + 1} {x^{(1)}
(t)} dt = 2b,\ k=2,3,\ldots,n-1. \label{Eq:after-integration}
\end{eqnarray}

By utilizing the Simpson numerical integration formula,
Eq.(\ref{Eq:after-integration}) can be expressed as
\begin{eqnarray}\label{Eq:substitution-z(k)}
&&x^{(1)} (k + 1) - x^{(1)} (k - 1) + a\frac{{x^{(1)} (k - 1) +
4x^{(1)} (k) + x^{(1)} (k + 1)}}{3} = 2b, \nonumber\\
&&\hspace{7.5cm} k=2,3,\ldots,n-1.
\end{eqnarray}

Eq.(\ref{Eq:substitution-z(k)}) turns to be
\begin{eqnarray}\label{Eq:after-substitution}
&&\left( {a + 3} \right)x^{(1)} (k + 1) + 4a x^{(1)} (k) + \left( {a
- 3} \right)x^{(1)} (k - 1) - 6b = 0,\nonumber\\
&&\hspace{6.8cm} k=2,3,\ldots,n-1.
\end{eqnarray}

It follows from Eq.(\ref{Eq:after-substitution}) that
\begin{eqnarray}
&&x^{(1)} (k + 1) - w x^{(1)} (k) = \frac{{a - 3}}{{w\left( {a + 3}
\right)}}\left[ {x^{(1)} (k) - w x^{(1)} (k - 1)} \right] +
\frac{{6b}}{{a + 3}},\nonumber\\
&&\hspace{7.5cm} k=2,3,\ldots,n-1, \label{Eq:x^{(1)}(k+1)-w
x^{(1)}(k)}
\end{eqnarray}
where $w = \frac{{\sqrt {3a^2  + 9}  - 2a}}{{a + 3}}$.

Iterating Eq.(\ref{Eq:x^{(1)}(k+1)-w x^{(1)}(k)}) by itself, we have
that
\begin{eqnarray}
&& x^{(1)} (k + 1) - w x^{(1)} (k) \nonumber \\
&&\hspace{7mm} \quad  = \frac{{a - 3}}{{w\left( {a + 3} \right)}}
                \left\{ {\frac{{a - 3}}{{w\left( {a + 3} \right)}}
                 \left[ {x^{(1)} (k - 1) - w x^{(1)} (k - 2)} \right]
                 + \frac{{6b}}{{a + 3}}} \right\} \nonumber\\
&&\hspace{15mm}+ \frac{{6b}}{{a + 3}} \nonumber\\
&&\hspace{7mm}\quad  = \left( {\frac{{a - 3}}{{w\left( {a + 3}
\right)}}} \right)^2
            \left[ {x^{(1)} (k - 1) - w x^{(1)} (k - 2)} \right]\nonumber\\
&&\hspace{15mm}
            + \frac{{6b}}{{a + 3}}\sum\limits_{i = 0}^1
            {\left( {\frac{{a - 3}}{{w\left( {a + 3} \right)}}} \right)} ^i \nonumber \\
&&\hspace{7mm} \quad  = \left( {\frac{{a - 3}}{{w\left( {a + 3}
\right)}}} \right)^{k - 1}
          \left[ {x^{(1)} (2) - w x^{(1)} (1)} \right]
           + \frac{{6b}}{{a + 3}}\sum\limits_{i = 0}^{k - 2}
           {\left( {\frac{{a - 3}}{{w\left( {a + 3} \right)}}} \right)^i }  \nonumber\\
&&\hspace{7mm} \quad  = \lambda ^{k - 1} \left[ {x^{(1)} (2) -
                 w x^{(1)} (1)} \right] + \mu \frac{{1 - \lambda ^{k - 1} }}
                 {{1 -\lambda }},  \label{Eq:Iterating-by-itself}
\end{eqnarray}
where $\lambda  = \frac{{a - 3}}{{w\left( {a + 3} \right)}}$, $\mu =
\frac{{6b}}{{a + 3}}$.

Note that
\begin{eqnarray}
&& x^{(1)} (k + 1) - w^{k - 1}x^{(1)} (2) \nonumber \\
&&\hspace{5mm} \quad  = \sum\limits_{j = 0}^{k - 2} {w^j }
           \left[ {x^{(1)} (k - j + 1) - w x^{(1)} (k - j)} \right] \nonumber\\
&&\hspace{5mm} \quad  = \sum\limits_{j = 0}^{k - 2} {w^j }
          \left\{ {\lambda ^{k - j - 1} \left[ {x^{(1)} (2) - w x^{(1)} (1)} \right]
           + \mu \frac{{1 - \lambda ^{k - j - 1} }}{{1 - \lambda }}} \right\} \nonumber\\
&&\hspace{5mm} \quad  = \sum\limits_{j = 0}^{k - 2} {w^j \lambda
                 ^{k - j - 1} \left[ {x^{(1)} (2) - w x^{(1)} (1)} \right]}  +
                     \frac{\mu }{{1 - \lambda }}\sum\limits_{j = 0}^{k - 2} {w^j \left(
                 {1 - \lambda ^{k - j - 1} } \right)}.\nonumber\\  \label{Eq:Note-that}
\end{eqnarray}

We have the 1-AGO series $\hat X^{(1)}$ of discrete {$\rm
GM_{SD}$}(1,1) is
\begin{eqnarray}
&&\hat x^{(1)} (k + 1) = w^{k - 1} x^{(1)} (2) + \left[ {x^{(1)} (2)
- w x^{(1)} (1)} \right]\sum\limits_{j = 0}^{k - 2} {w^j \lambda ^{k
- j - 1} } \nonumber\\
&&\hspace{25mm} + \frac{\mu }{{1 - \lambda }}\sum\limits_{j = 0}^{k
- 2} {w^j \left( {1 - \lambda ^{k - j - 1} } \right)},
 k=2,3,\ldots,n-1.  \label{Eq:responsed-values-of-x1(k+1)}
\end{eqnarray}

Apply first-order inverse accumulation operation to obtain the
simulation and forecasting value
\begin{eqnarray}
&&\hat x^{(0)} \left( {k + 1} \right) = \hat x^{(1)} \left( {k + 1}
\right) - \hat x^{(1)} \left( k \right) \nonumber\\
&&\hspace{19mm} = w^{k - 2} \left( {w - 1} \right) x^{(1)} (2)\nonumber\\
&&\hspace{23mm} + \left[ {x^{(1)} (2) - w x^{(1)} (1)} \right]\left[
{\left( {\lambda - 1} \right)\sum\limits_{j = 0}^{k - 2} {w^j
\lambda ^{k - 2 - j} } +
w^{k - 2} } \right] \nonumber\\
&&\hspace{23mm}+ \mu \sum\limits_{j = 0}^{k - 2} {w^j \lambda ^{k -
2 - j} },\ k=1,2,\ldots,n-1.
\label{Eq:first-order-inverse-accumulation-operation}
\end{eqnarray}

Now the discrete ${\rm GM_{SD}}$(1,1) model has been constructed,
and the whole modeling procedure is analyzed.
\subsection{Parameters estimation of the discrete ${\rm GM_{SD}}$(1,1) model}
\label{susbsec:estination}

From the definition of 1-AGO, we have that
\begin{eqnarray}
x^{(1)} (k + 1) - x^{(1)} (k - 1) = x^{(0)} (k + 1) + x^{(0)} (k).
\nonumber
\end{eqnarray}

By the Simpson numerical integration formula, the background value
of $X^{(1)}$ is defined as
\begin{eqnarray}
z^{\left( 1 \right)} (k) = \frac{{x^{(1)} (k - 1) + 4x^{(1)} (k) +
x^{(1)} (k + 1)}}{3}. \nonumber
\end{eqnarray}

Thus the Eq.(\ref{Eq:substitution-z(k)}) can be rewritten as below
\begin{eqnarray}
x^{(0)} (k + 1) + x^{(0)} (k) + a z^{\left( 1 \right)} (k) = 2b.
\label{Eq:rewriten-substitution-z(k)}
\end{eqnarray}

Employing the least squares estimation method, from
Eq.(\ref{Eq:rewriten-substitution-z(k)}) by considering
$k=2,3,\ldots,n-1$, the model parameters $a$ and $b$ can be given as
\begin{eqnarray}\label{Eq:least-squares-estimation-GMSD}
\left( {\begin{array}{*{20}c}
   {\hat a}  \\
   {\hat b}  \\
\end{array}} \right) = \left( {\boldsymbol{B}^T \boldsymbol{B}} \right)^{ - 1} \boldsymbol{B}^T \boldsymbol{Y},
\end{eqnarray}
where $\boldsymbol{B}$ and $\boldsymbol{Y}$ are defined as follows
\begin{eqnarray}
\boldsymbol{B} = \left( {\begin{array}{*{20}c}
   { - \frac{{x^{(1)} (1) + 4x^{(1)} (2) + x^{(1)} (3)}}{6}} &{}& 1  \\
   { - \frac{{x^{(1)} (2) + 4x^{(1)} (3) + x^{(1)} (4)}}{6}} &{}& 1  \\
                           \vdots                        &{}&  \vdots   \\
   { - \frac{{x^{(1)} (n - 2) + 4x^{(1)} (n - 1) + x^{(1)} (n)}}{6}} &{}& 1  \\
\end{array}} \right),\quad
\boldsymbol{Y} = \left( {\begin{array}{*{20}c}
   {\frac{{x^{(0)} (2) + x^{(0)} (3)}}{2}}  \\
   {\frac{{x^{(0)} (3) + x^{(0)} (4)}}{2}}  \\
    \vdots   \\
   {\frac{{x^{(0)} (n - 1) + x^{(0)} (n)}}{2}}  \\
\end{array}} \right). \nonumber
\end{eqnarray}
\subsection{Difference between ${\rm GM_{SC}}$(1,1) and ${\rm GM_{SD}}$(1,1) models}
\label{subsec:Difference}

This subsection discusses the difference between the continuous
${\rm GM_{SC}}$(1,1) model and the discrete ${\rm GM_{SD}}$(1,1)
model. In the paper of He and Wang \cite{He2011ConstructingJQE}, the
time response function is expressed by
\begin{eqnarray}
\hat x^{(1)} (k + 1) = \left( {x^{(0)} (1) - \frac{b}{a}} \right)e^{
- ak}  + \frac{b}{a},\ k=1,2,\ldots,n-1,
\label{Eq:response-function-HeandWang}
\end{eqnarray}
and the restored values of $\hat x^{(0)} \left( {k + 1} \right)$ is
given by
\begin{eqnarray}
&&\hat x^{(0)} (k + 1) = \hat x^{(1)} \left( {k + 1} \right) - \hat
x^{(1)} \left( k \right) = \frac{{e^a  - 1}}{a}\left( {b - ax^{(0)}
(1)} \right)e^{ - ak},\nonumber\\
&&\hspace{7cm} k=1,2,\ldots,n-1.
\label{Eq:restored-values-HeandWang}
\end{eqnarray}

They are the same as the ones of the classical continuous GM(1,1)
model provided in Section \ref{sec:the-basis}. The system parameters
$a$ and $b$ in Eqs.(\ref{Eq:response-function-HeandWang}) and
(\ref{Eq:restored-values-HeandWang}) are derived from the least
squares estimation solution of the
Eq.(\ref{Eq:rewriten-substitution-z(k)}). Obviously, the function
(\ref{Eq:response-function-HeandWang}) must coincidence with the
difference Eq.(\ref{Eq:rewriten-substitution-z(k)}), otherwise the
continuous ${\rm GM_{SC}}$(1,1) model will not be accurate.
Substituting the expression (\ref{Eq:response-function-HeandWang})
into the Eq.(\ref{Eq:rewriten-substitution-z(k)}), the left side of
Eq.(\ref{Eq:rewriten-substitution-z(k)}) turns to be
\begin{eqnarray}
&&L(t) = x^{(0)} (k + 1) + x^{(0)} (k) + a z^{\left( 1 \right)} (k) \nonumber\\
&&\hspace{7.5mm}= x^{(1)} (k + 1) - x^{(1)} (k) + a\frac{{x^{(1)} (k
   - 1) + 4x^{(1)} (k) + x^{(1)} (k +
        1)}}{3} \nonumber\\
&&\hspace{7.5mm} = \frac{1}{3}\left[ {\left( {a + 3} \right)x^{(1)}
                (k + 1) + 4ax^{(1)} (k) + \left( {a - 3} \right)x^{(1)} (k - 1)}
        \right] \nonumber \\
&&\hspace{7.5mm} = \frac{1}{3}\left( {x^{(0)} (1) - \frac{b}{a}}
\right)e^{ - ak}
          \left[ {\left( {a + 3} \right) + 4ae^a  + \left( {a - 3} \right)e^{2a} } \right]  \nonumber\\
&&\hspace{8.5mm} \quad  + \frac{b}{{3a}}\left[ {\left( {a + 3}
\right)
              + 4a + \left( {a - 3} \right)} \right]  \nonumber\\
&&\hspace{7.5mm}  = \frac{1}{3}\left( {x^{(0)} (1) - \frac{b}{a}}
               \right)e^{ - ak} \left[ {\left( {a + 3} \right) + 4a e^a  + \left( {a
             - 3} \right)e^{2a} } \right] + 2b. \label{Eq:value-of-left}
\end{eqnarray}

The right side of Eq.(\ref{Eq:rewriten-substitution-z(k)}) is
\begin{eqnarray}\label{Eq:value-of-right}
R(t) = 2b.
\end{eqnarray}

\begin{figure}[!htbp]
\centering\centerline{\includegraphics[height=6cm,width=7cm]{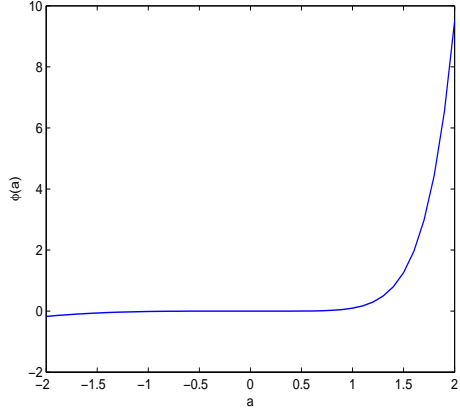}}
\caption{Function $\phi(a)$ for different values of $a$}
 \label{fig:function-phi(a)}
\end{figure}

Let $\phi(a) = \left( {a + 3} \right) + 4ae^a  + \left( {a - 3}
\right)e^{2a}$, we obtain the following numerical result displayed
in Table \ref{table:function-phi(a)} and Fig.
\ref{fig:function-phi(a)}.
 \newpage
\begin{table}[!htbp]
\caption{Computation results of function $\phi(a)$ under different
values of $a$}
 \label{table:function-phi(a)}
 \centering
 \footnotesize
\begin{tabular}{llllllllllllll}
\hline
 &$a$   &$\phi(a)$                &\quad &$a$   &$\phi(a)$                 &\quad &$a$     &$\phi(a)$\\
 \hline
  &0.00  & 0                      &\quad &0.35   &2.4989$\times 10^{-4}$ &\quad &0.70    &0.0115         \\
  &0.05  & 1.0952$\times 10^{-8}$ &\quad &0.40  &5.1310$\times 10^{-4}$ &\quad &0.75    &0.0172         \\
  &0.10  & 3.6857$\times 10^{-7}$ &\quad &0.45  &9.7400$\times 10^{-4}$ &\quad &0.80    &0.0251         \\
  &0.15  & 2.9440$\times 10^{-6}$ &\quad &0.50  &0.0017                 &\quad &0.85    &0.0358         \\
  &0.20  & 1.3053$\times 10^{-5}$ &\quad &0.55  &0.0029                 &\quad &0.90    &0.0503        \\
  &0.25  & 4.1922$\times 10^{-5}$ &\quad &0.60  &0.0048                 &\quad &0.95    &0.0696         \\
  &0.30  & 1.0981$\times 10^{-4}$ &\quad &0.65  &0.0076                 &\quad &1.00     &0.0950         \\
\hline
\end{tabular}
\end{table}

One checks easily that when $\left| a \right|$ approximately to
zero, the first term of Eq.(\ref{Eq:value-of-left}) is approximately
to zero. In this situation, we can say $L(t) = R(t)$. However, when
$\left| a \right|$ is large $(a \ge 1.5)$, the errors $L(t) - R(t)$
will be quite large. That implies the function
(\ref{Eq:response-function-HeandWang}) will not coincidence with the
difference Eq.(\ref{Eq:rewriten-substitution-z(k)}), and the
continuous ${\rm GM_{SC}}$(1,1) model may not be accurate. On the
other hand, the discrete function
(\ref{Eq:responsed-values-of-x1(k+1)}) is exactly the solution of
the difference Eq.(\ref{Eq:rewriten-substitution-z(k)}). This means
the performance of the discrete ${\rm GM_{SD}}$(1,1) model is not
limited to the value of system parameters.

In the above analysis, the difference between the continuous ${\rm
GM_{SC}}$(1,1) model and the discrete ${\rm GM_{SD}}$(1,1) model is
that the modelling accuracy of the former depends on system
parameters' value, while the later does not. This is the advantage
of the discrete model compared to the continuous one.
\subsection{Unbiased property of the discrete ${\rm GM_{SD}}$(1,1) model}
\label{subsec:unbiased}

This subsection proves the discrete ${\rm GM_{SD}}$(1,1) model is
unbiased to simulate the homogeneous exponent sequence. Set the
sequence is $\left\{ {rq^k ,k = 1,2, \ldots ,n} \right\}$, then the
following original sequence as $X^{(0)}  = \left( {rq,rq^2 , \ldots
,rq^n } \right)$. One checks easily that
\begin{eqnarray}
x^{(1)} (k) = \sum\limits_{i = 1}^k {x^{(0)} (i)}  = \frac{{rq\left(
{1 - q^k } \right)}}{{1 - q}},k = 1,2, \ldots ,n. \nonumber
\end{eqnarray}

The 1-AGO of $X^{(0)}$ is given by
\begin{eqnarray}
X^{(1)}  = \left\{ {r q,\frac{{rq\left( {1 - q^2 } \right)}}{{1 -
q}},\frac{{rq\left( {1 - q^3 } \right)}}{{1 - q}}, \ldots
,\frac{{rq\left( {1 - q^n } \right)}}{{1 - q}}} \right\}. \nonumber
\end{eqnarray}

Substituting these values into the matrix $\boldsymbol{B}$ and
$\boldsymbol{Y}$, it follows that
\begin{eqnarray}
&&\boldsymbol{B}= \left( {\begin{array}{*{20}c}
   { - \frac{{6rq - rq^2 \left( {1 + 4q + q^2 } \right)}}{{6\left( {1 - q} \right)}}} & 1  \\
   { - \frac{{6rq - rq^3 \left( {1 + 4q + q^2 } \right)}}{{6\left( {1 - q} \right)}}} & 1  \\
    \vdots  &  \vdots   \\
   { - \frac{{6rq - rq^{n - 1} \left( {1 + 4q + q^2 } \right)}}{{6\left( {1 - q} \right)}}} & 1  \\
\end{array}} \right), \quad \boldsymbol{Y} = \left( {\begin{array}{*{20}c}
   {\frac{{rq^2 \left( {1 + q} \right)}}{2}}  \\
   {\frac{{rq^3 \left( {1 + q} \right)}}{2}}  \\
    \vdots   \\
   {\frac{{rq^{n - 1} \left( {1 + q} \right)}}{2}}  \\
\end{array}} \right). \nonumber
\end{eqnarray}

After some calculations, we known that
\begin{eqnarray}
\left( {\begin{array}{*{20}c}
   a  \\
   b  \\
\end{array}} \right) = \left( {\boldsymbol{B}^T \boldsymbol{B}} \right)^{ - 1} \boldsymbol{B}^T \boldsymbol{Y} = \left( {\begin{array}{*{20}c}
   {\frac{{3\left( {1 - q^2 } \right)}}{{1 + 4q + q^2 }}}  \\ [3pt]
   {\frac{{3rq\left( {1 + q} \right)}}{{1 + 4q + q^2 }}}  \\
\end{array}} \right).  \label{Eq:a-and-b-values}
\end{eqnarray}

From Eq.(\ref{Eq:a-and-b-values}), we can easily obtain
\begin{eqnarray}
w = q, \quad \lambda  =  - \frac{{q + 2}}{{2q + 1}},\quad \mu  =
\frac{{3rq\left( {1 + q} \right)}}{{2q + 1}}. \nonumber
\end{eqnarray}

Substituting the three values into the
Eq.(\ref{Eq:first-order-inverse-accumulation-operation}), it yields
that
\begin{eqnarray}
&&\hspace{-1cm}\hat x^{(0)} (k + 1) = q^{k - 2} \frac{{rq\left( {1 -
q^2 } \right)}}{{1 -
q}}\left( {q - 1} \right)\nonumber\\
 &&\hspace{9mm}+ \left( {\frac{{rq\left( {1 - q^2 }
\right)}}{{1 - q}} - qrq} \right)\left( { - \frac{{3q + 3}}{{2q +
1}}\sum\limits_{j =
0}^{k - 2} {w^j \lambda ^{k - 2 - j} }  + q^{k - 2} } \right)\nonumber\\
 &&\hspace{9mm}+
\frac{{3qr\left( {q + 1} \right)}}{{2q + 1}}\sum\limits_{j = 0}^{k -
2} {w^j \lambda ^{k - 2 - j} }\nonumber\\
&&\hspace{8mm} =  - rq^{k - 1}  + rq^{k + 1}  + rq\left( { -
\frac{{3\left( {q + 1} \right)}}{{2q + 1}}\sum\limits_{j = 0}^{k -
2} {w^j \lambda ^{k - 2 - j} }  + q^{k - 2} } \right) \nonumber\\
 &&\hspace{9mm} +
\frac{{3qr\left( {q + 1} \right)}}{{2q + 1}}\sum\limits_{j = 0}^{k -
2} {w^j \lambda ^{k - 2
- j} }\nonumber \\
&&\hspace{8mm} = rq^{k + 1}  = x^{(0)} (k + 1).
\label{Eq:verify-is-succeesed}
\end{eqnarray}

Eq.(\ref{Eq:verify-is-succeesed}) indicates that the homogeneous
exponent simulative unbiased property is met.
{
\subsection{Modelling evaluation criteria}
 \label{subsec:criteria}

 To examine the prediction accuracy of the ${\rm GM_{SD}}$(1,1) model,
the absolute percentage error (APE) and the mean absolute percentage
error (MAPE) are adopted in this paper. They are defined as follows
\begin{eqnarray}
&&{\rm APE}(k) = \left| {\frac{{ x^{(0)} (k) - \hat x^{(0)}
(k)}}{{x^{(0)} (k)}}} \right| \times 100\%,\ k=2,3,\ldots,n,
 \label{Eq:APE}\\
 &&{\rm MAPE}= \frac{1}{{m - \ell+ 1}}\sum\limits_{k = \ell}^m {\left| {\frac{{ x^{(0)} (k) - \hat x^{(0)}
(k)}}{{x^{(0)} (k)}}} \right| \times 100\% },\ m\le n.
  \label{Eq:MAPE}
\end{eqnarray}

From Eq.(\ref{Eq:APE}), ${\rm APE}(k),\ k=2,3,\ldots,\nu$ is
referred to as the absolute simulation percentage error at time $k$,
while ${\rm APE}(k),\ k=\nu+1,\nu+2,\ldots,n$ is referred to as the
absolute prediction percentage error at time $k$. Further, when
$\ell=2, m=\nu$, the MAPE is the mean absolute simulation percentage
error termed ${\rm MAPE_{\rm simu}}$, when $\ell=\nu+1, m=n$, the
MAPE is the mean absolute prediction percentage error termed ${\rm
MAPE_{\rm pred}}$, and when $\ell=2, m=n$, the MAPE is the overall
mean absolute percentage error termed ${\rm MAPE_{\rm over}}$.  }
{
\section{Validation of the ${\rm GM_{SD}}$(1,1) model}
  \label{sec:validation}

This section provides some numerical examples to validate the
accuracy of the ${\rm GM_{SD}}$(1,1) model compared to the classical
GM(1,1) model, the DGM(1,1) model and the ${\rm GM_{SC}}$(1,1)
model.

\subsection{Validation of ${\rm GM_{SD}}$(1,1) and ${\rm GM_{SC}}$(1,1) models}

This subsection provides an example to verify} the accuracy of the
${\rm GM_{SD}}$(1,1) model and the ${\rm GM_{SC}}$(1,1) model to
simulate and predict the homogeneous exponent sequence. Let $x^{(0)}
(k) = rq^k ,\ k = 1,2, \ldots ,12,\ r > 0$, where parameter $r$ is
randomly generated in [1, 15] by the discrete uniform distribution,
and parameter $q$ is given in the intervals [0.1, 5.0] by the step
0.01. We define the following notation in the sequel
\begin{eqnarray}
&&\varepsilon = \left| {\hat a - a} \right| + \left| {\hat b - b}
\right|, \label{Eq:epsilon}
\end{eqnarray}
where $\hat{a}$ and $\hat{b}$ are the estimated parameters of ${\rm
GM_{SD}}$(1,1) and ${\rm GM_{SC}}$(1,1) models, and parameters $a$
and $b$ are the provided determined of Eq.(\ref{Eq:a-and-b-values}).

\begin{figure}[!htbp]
\centering\centerline{\includegraphics[height=5.6cm,width=8.cm]{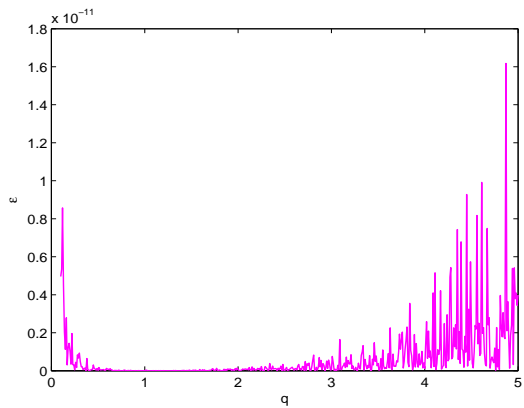}}
\caption{The values of $\varepsilon$ for different values of $q$ and
$r$}
 \label{figure:unbaised-values}
\end{figure}

{Employing the above specific parameters, the graphs are depicted in
Fig. \ref{figure:unbaised-values}. It can be seen in Fig.
\ref{figure:unbaised-values} that} the maximum $\varepsilon$ is only
$1.6172 \times 10^{ - 11}$ which is obvious a truncation error by
computer.

{Further, fixed parameters $r=0.05$ and $q=2.25$, results of the
${\rm GM_{SC}}$(1,1) and the ${\rm GM_{SD}}$(1,1) models are listed
in Table \ref{table:unbaised-values}. We observe from Table
\ref{table:unbaised-values} that the maximum absolute simulation
percentage error of ${\rm GM_{SC}}$(1,1) and ${\rm GM_{SD}}$(1,1)
models are, respectively, 1.0397\% and $0.1559\times 10^{-12}$\%,
and the maximum absolute prediction percentage error of ${\rm
GM_{SC}}$(1,1) and ${\rm GM_{SD}}$(1,1) models are, respectively,
2.1037\% and $0.0945\times 10^{-12}$\%.} Obviously, the APE of the
${\rm GM_{SD}}$(1,1) model is caused by the round-off error of
computer, while the APE of the ${\rm GM_{SC}}$(1,1) model is caused
by its inconsistency.
\begin{table}[!htbp]
\caption{The predictive and error values with $r=0.05$ and $q=2.25$}
 \label{table:unbaised-values}
 \centering
 \footnotesize
\begin{tabular}{ccccccccc}
\hline
$k$   & actual values  &\multicolumn{2}{c}{${\rm GM_{SC}}$(1,1) model}    &\quad  &\multicolumn{2}{c}{${\rm GM_{SD}}$(1,1) model}    \\
                                    \cline{3-4}                         \cline{6-7}
        &             & values    &APE($k$)\%  &\quad  & values          &APE($k$)\% \\
\noalign{\smallskip}\hline\noalign{\smallskip}
1       &0.1125       & 0.1125             &0.0000    &\quad  &0.1125                     &0.0000            \\
2       &0.2531       & 0.2531             &0.0000    &\quad  &0.2531                     &0.0000            \\
3       &0.5695       & 0.5667             &0.5034    &\quad  &0.5695                     &\underline{0.1559$\times 10^{-12}$}         \\
4       &1.2814       & 1.2727             &0.6825    &\quad  &1.2814                     &0.0520$\times 10^{-12}$       \\
5       &2.8833       & 2.8584             &0.8613    &\quad  &2.8833                     &0.0462$\times 10^{-12}$      \\
6       &6.4873       & 6.4199             &\underline{1.0397}    &\quad  &6.4873     &0.0274$\times 10^{-12}$ \\[4pt]
7       &14.5965      &14.4187             &1.2178    &\quad  &14.5965                    &0.0122$\times 10^{-12}$  \\
8       &32.8420      &32.3837             &1.3956    &\quad  &32.8420                    &0.0216$\times 10^{-12}$ \\
9       &73.8946      &72.7321             &1.5731    &\quad  &73.8946                    &0.0192$\times 10^{-12}$ \\
10      &166.2628     &163.3528            &1.7503    &\quad  &166.2628                   &0.0342$\times 10^{-12}$ \\
11      &374.0914     &366.8821            &1.9271    &\quad  &374.0914                   &0.0456$\times 10^{-12}$ \\
12      &841.7056     &823.9990            &\underline{2.1037}    &\quad  &841.7056      &\underline{0.0945$\times 10^{-12}$} \\[2pt]
\multicolumn{3}{l}{${\rm MAPE_{\rm simu}}$} & 0.7717 &  &   &{\bf \underline {1.5565$\times \bf {10^{-13}}$ }}\\
\multicolumn{3}{l}{${\rm MAPE_{\rm pred}}$} & 1.6613 &  &   &{\bf \underline {3.7893$\times \bf {10^{-14}}$ }}\\
\multicolumn{3}{l}{${\rm MAPE_{\rm over}}$} &1.3055  &  &   &{\bf \underline {5.0888$\times \bf {10^{-14}}$ }}\\
\hline
\end{tabular}
\end{table}

{
\subsection{Validation of ${\rm GM_{SD}}$(1,1) and other grey models}

This subsection further illustrates the advantage of the ${\rm
GM_{SD}}$(1,1) model by using some real cases. We consider the
numerical example from the paper \cite{Ding2018forecastingE} to
predict total electricity consumption in China during 2005-2014.
Data from 2005 to 2011 are applied to develop different grey models,
while data from 2012 to 2014 are applied to test. Results are
presented in Table \ref{tab:validation-electricity} showing that the
${\rm GM_{SD}}$(1,1) model outperforms the other grey models in this
example.  }

\begin{table}[!htbp]

\caption{Results of GM(1,1), DGM(1,1), ${\rm GM_{SC}}$(1,1) and
${\rm GM_{SD}}$(1,1) models }
 \label{tab:validation-electricity}
 \centering
 \footnotesize
 {
 \begin{tabular}{cccccccccccccc}
 \hline
Year    &Data    &  GM(1,1)     & DGM(1,1)    & ${\rm GM_{SC}}$(1,1)  &${\rm GM_{SD}}$(1,1) \\
\hline
2005    &24940.3 &24940.3000    &24940.3000   &24940.3000             &24940.3000      \\
2006    &28588.0 &28678.0326    &28701.0256   &28588.0000             &28588.0000      \\
2007    &32711.8 &31558.7319    &31586.3461   &32127.3503             &32080.3602     \\
2008    &34541.4 &34728.7965    &34761.7284   &34825.5851             &34472.6239    \\
2009    &37032.2 &38217.2931    &38256.3326   &38190.8687             &38490.1274   \\
2010    &41932.5 &42056.2081    &42102.2502   &41881.3482             &41543.9697     \\
2011    &47000.9 &46280.7409    &46334.7987   &45928.4479             &46203.8856   \\[4pt]
2012    &49762.6 &50929.6267    &50992.8462   &50366.6289             &50042.7998    \\
2013    &54203.4 &56045.4916    &56119.1683   &55233.6827             &55485.5079    \\
2014    &56383.7 &61675.2435    &61760.8406   &60571.0519             &60258.6361   \\[4pt]
\multicolumn{2}{l}{${\rm MAPE_{\rm simu}}$} &{\underline{\bf 1.5675\%}}  &1.5994\% &1.6293\%    &1.7387\%\\
\multicolumn{2}{l}{${\rm MAPE_{\rm pred}}$} &5.0428\%  &5.1811\% &3.5137\%    &{\underline{\bf 3.2669\%}}\\
\multicolumn{2}{l}{${\rm MAPE_{\rm over}}$} &2.7260\%  &2.7933\% &2.3360\%    &{\underline{\bf 2.3118\%}}\\
\hline
\end{tabular} }
\end{table}


\section{Applications}
\label{sec:applications-example}

In this section, the discrete ${\rm GM_{SD}}$(1,1) model is used to
predict the Gross Domestic Product (GDP) and the freightage {of
Lanzhou.}
\subsection{Forecasting the Gross Domestic Product of {Lanzhou}}
\label{sec:GDP}

\begin{figure}[!htbp]
\centering
\begin{minipage}[c]{0.5\textwidth}
\centering
\includegraphics[height=5.4cm,width=6.2cm]{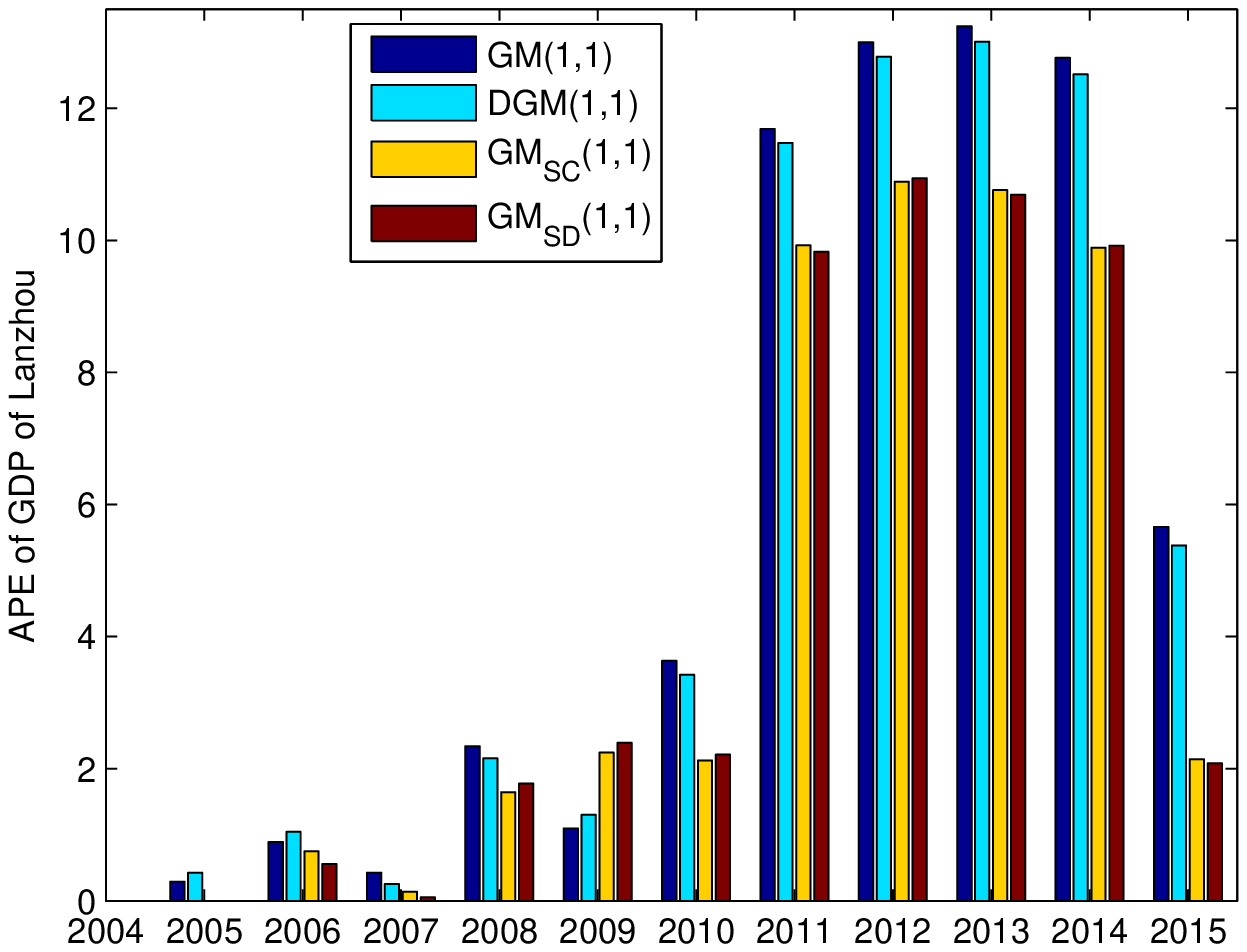}
\end{minipage}%
\begin{minipage}[c]{0.5\textwidth}
\centering
\includegraphics[height=5.4cm,width=6.2cm]{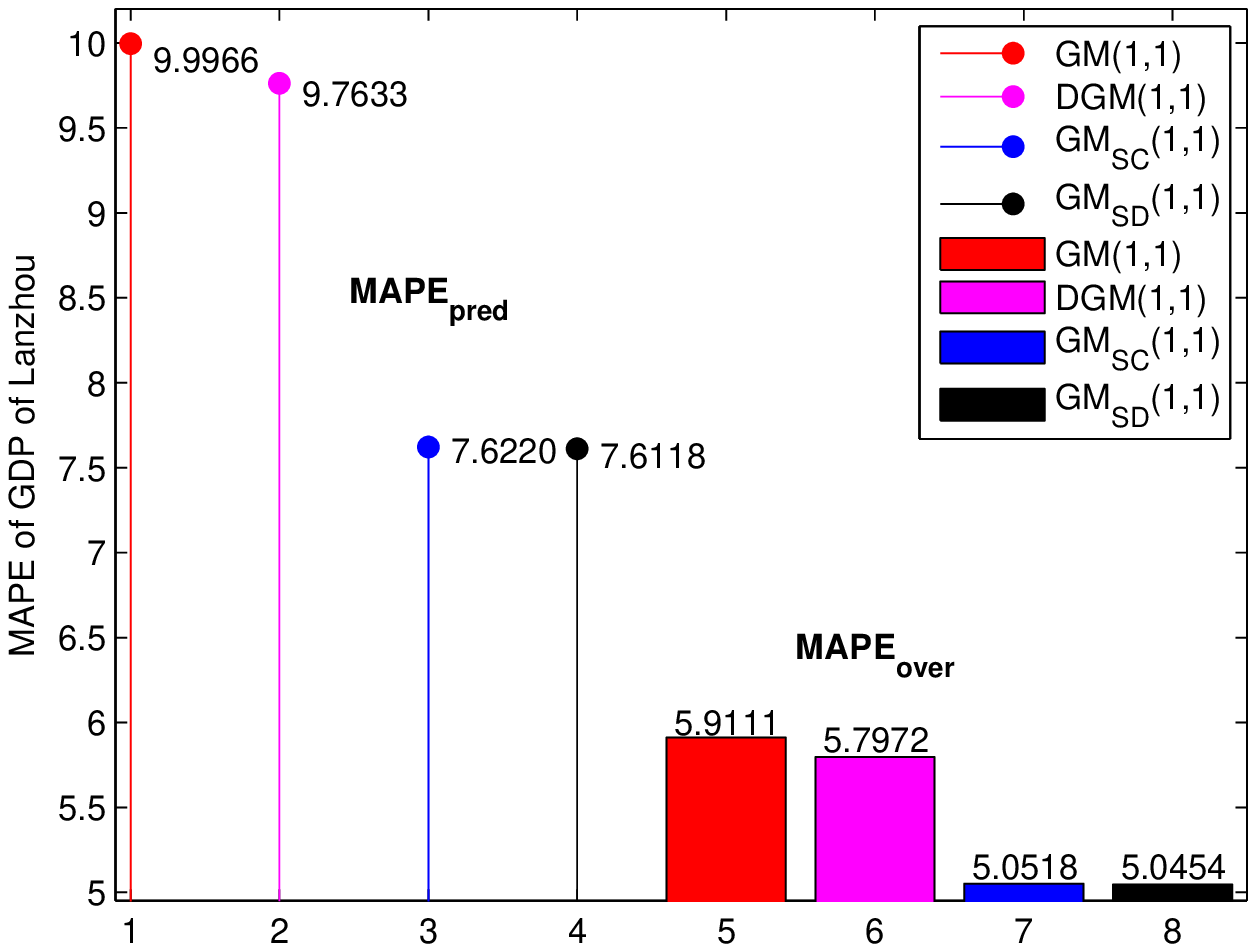}
\end{minipage}
\caption{Percentage errors among four models for GDP}
\label{fig:error-gdp}
\end{figure}

Raw data of Lanzhou was collected from the website of the National
Bureau of Statistics of China. The total Gross Domestic Product is
measured in hundred million RMB. These real data from 2004 to 2009
are applied to build the prediction models, and the ones from 2010
to 2015 are applied for validation. {The simulation and prediction
results are listed in Table \ref{table:prediction-GDP}, while the
errors are listed in Table \ref{table:err-GDP}, and Fig.
\ref{fig:error-gdp}.

From Tables \ref{table:prediction-GDP} and \ref{table:err-GDP}, and
Fig. \ref{fig:error-gdp}  } that four grey models have successfully
caught the trend of the GDP. The ${\rm GM_{SD}}$(1,1) model for {the
mean absolute prediction percentage error and the overall mean
absolute percentage error} are 7.6118\% and 5.0454\%, respectively,
which have the smallest errors among four grey models. Fig.
\ref{fig:error-gdp} also indicates that the accuracy of ${\rm
GM_{SD}}$(1,1) model is the best, and the GM(1,1) model is the
worst.
\begin{table}[!htbp]
 \caption{Simulation and prediction results of GDP of Lanzhou}
 \label{table:prediction-GDP}
\centering
 \footnotesize
\begin{tabular}{ccccccccccccccccc}
\hline
 Year    & Data        &{GM}(1,1)   &DGM(1,1)  &{${\rm GM_{SC}}$(1,1)}  &{${\rm GM_{SD}}$(1,1) }  \\
 \hline
 2004    &504.65       &504.6500    &504.6500  &504.6500                &504.6500                   \\
 2005    &567.04       &568.6831    &569.4678  &567.0400                &567.0400                   \\
 2006    &638.47       &644.1549    &645.1480  &643.2803                &642.0348                 \\
 2007    &732.76       &729.6429    &730.8858  &731.7331                &733.1665                 \\
 2008    &846.28       &826.4761    &828.0179  &832.3484                &831.2488                 \\
 2009    &926.00       &936.1605    &938.0585  &946.7986                &948.1637                 \\ [5pt]
 2010    &1100.40      &1060.4014   &1062.7230 &1076.9861               &1076.0331                \\
 2011    &1360.03      &1201.1307   &1203.9551 &1225.0746               &1226.3916                \\
 2012    &1563.80      &1360.5367   &1363.9563 &1393.5258               &1392.7242                \\
 2013    &1776.28      &1541.0980   &1545.2212 &1585.1395               &1586.4313                \\
 2014    &2000.94      &1745.6222   &1750.5754 &1803.1005               &1802.4595                \\
 2015    &2095.99      &1977.2894   &1983.2206 &2051.0319               &2052.3253                \\
\hline
\end{tabular}
\end{table}
\begin{table}[!htbp]
 \caption{Relative error values of GDP of Lanzhou by grey models (\%)}
 \label{table:err-GDP}
\centering
 \footnotesize
\begin{tabular}{ccccccccccccccccc}
\hline
 year    &{GM}(1,1)                 &DGM(1,1)  &{${\rm GM_{SC}}$(1,1)}  &{${\rm GM_{SD}}$(1,1) }  \\
 \hline
 2004    &0                         &0         &0                       &0    \\
 2005    &0.2898                    &0.4281    &0                       &0\\
 2006    &0.8904                    &1.0459    &0.7534                  &0.5583\\
 2007    &0.4254                    &0.2558    &0.1401                  &0.0555\\
 2008    &2.3401                    &2.1579    &1.6462                  &1.7762  \\
 2009    &1.0972                    &1.3022    &2.2461                  &2.3935 \\
 2010    &3.6349                    &3.4239    &2.1278                  &2.2144 \\
 2011    &11.6835                   &11.4758   &9.9230                  &9.8261 \\
 2012    &12.9980                   &12.7794   &10.8885                 &10.9397  \\
 2013    &13.2401                   &13.0080   &10.7607                 &10.6880  \\
 2014    &12.7599                   &12.5123   &9.8873                  &9.9194 \\
 2015    &5.6632                    &5.3802    &2.1450                  &2.0832  \\[5pt]
 {${\rm MAPE_{\rm simu}}$} &{\underline{\bf 1.0086}}  &1.0379 &1.1869    &1.1959\\
 {${\rm MAPE_{\rm pred}}$} &9.9966  &9.7633 &7.6220    &{\underline{\bf 7.6118}}\\
 {${\rm MAPE_{\rm over}}$} &5.9111  &5.7972 &5.0518    &{\underline{\bf 5.0454}}\\
\hline
\end{tabular}
\end{table}

\subsection{Forecasting the freightage of {Lanzhou}}
\label{subsec:freightage}

\begin{figure}[!htbp]
\centering
\begin{minipage}[c]{0.5\textwidth}
\centering
\includegraphics[height=5.4cm,width=6.2cm]{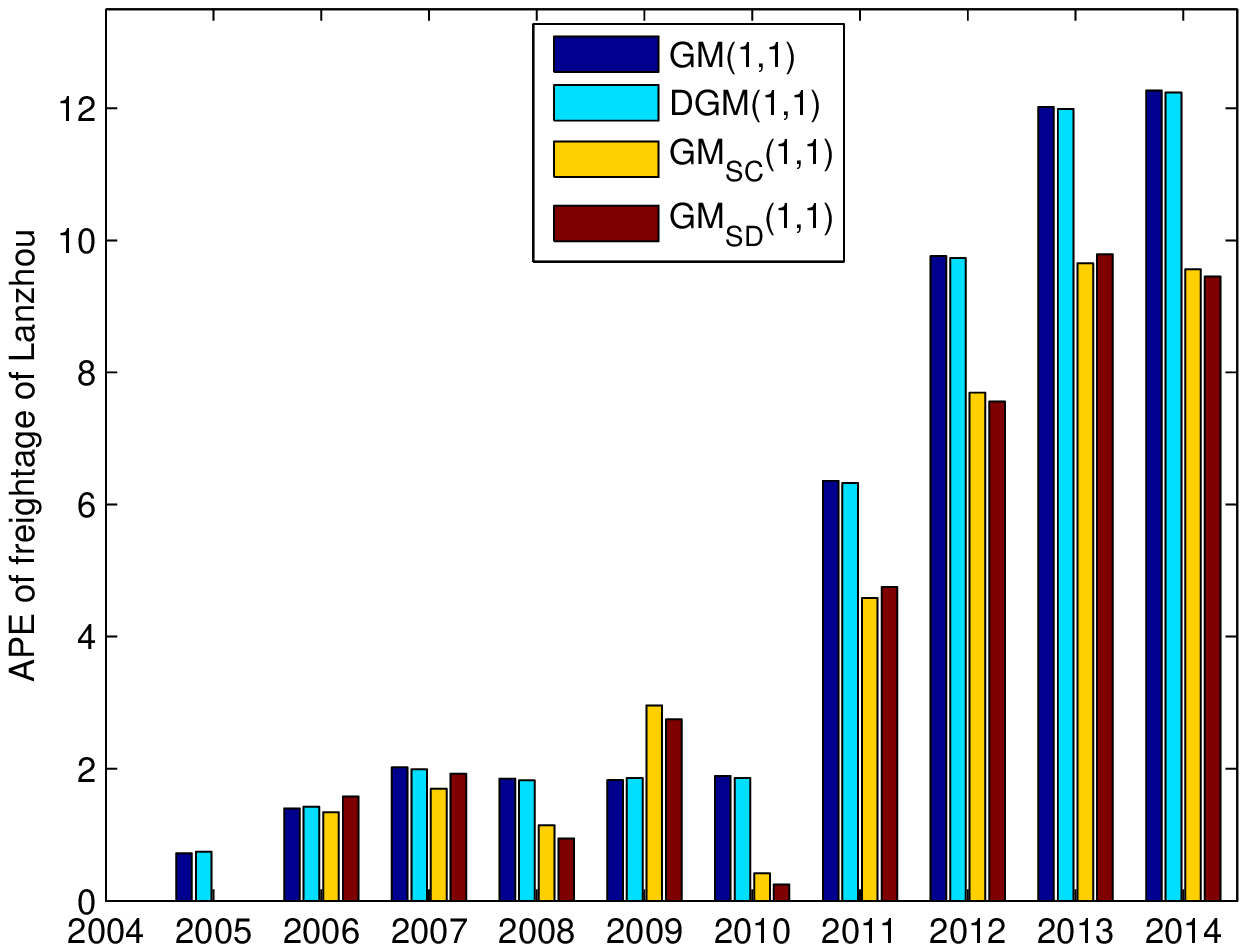}
\end{minipage}%
\begin{minipage}[c]{0.5\textwidth}
\centering
\includegraphics[height=5.4cm,width=6.2cm]{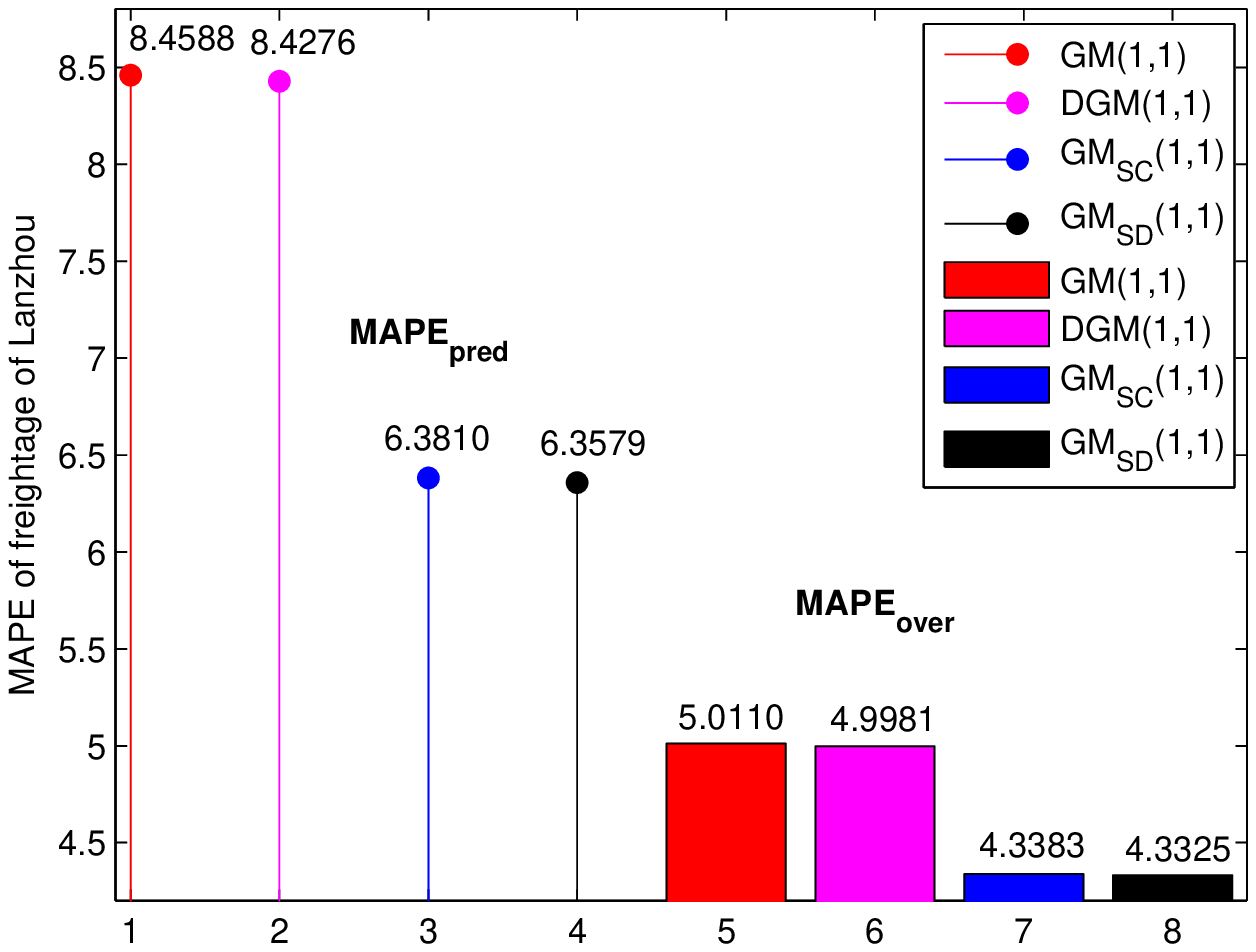}
\end{minipage}
\caption{Percentage errors among four models for freightage}
\label{fig:error-freightage}
\end{figure}

{The raw data of the freightage of Lanzhou was also collected from
the website of the National Bureau of Statistics of China. The total
freightage is measured in ten thousand tons. Similarly, we divided
these data into two groups, in which the first 6 samples are applied
to build the prediction models, and the left samples are used to
check and compare the forecasting results. The simulation and
prediction results are listed in Table
\ref{table:prediction-freightage}, while the errors are listed in
Table \ref{table:err-freightage} and Fig.
\ref{fig:error-freightage}.

We observe from Table \ref{table:err-freightage} and Fig.
\ref{fig:error-freightage} that the ${\rm MAPE_{\rm simu}}$, ${\rm
MAPE_{\rm pred}}$ and ${\rm MAPE_{\rm over}}$ of ${\rm
GM_{SD}}$(1,1) model are 1.8007\%, 6.3579\% and 4.3325\%,
respectively. ${\rm MAPE_{\rm simu}}$, ${\rm MAPE_{\rm pred}}$ and
${\rm MAPE_{\rm over}}$ of GM(1,1) are 1.5632\%, 8.4588\% and
5.0110\%, those of DGM(1,1) are 1.5687\%, 8.4276\% and 4.9981\%, and
those of ${\rm GM_{SC}}$(1,1) model are 1.8499\%, 6.3810\% and
4.3383\%, respectively.

Here the ${\rm GM_{SD}}$(1,1) model for the mean absolute prediction
percentage error and the overall mean absolute percentage error are
the smallest errors among four grey models. This also indicates that
the accuracy of ${\rm GM_{SD}}$(1,1) model is the best, the accuracy
of ${\rm GM_{SC}}$(1,1) model are inferior to GM(1,1) model and
DGM(1,1) model and the GM(1,1) model is the worst to predict the
freightage of Lanzhou. }
\begin{table}[!htbp]
 \caption{Simulation and prediction results of freightage of Lanzhou}
 \label{table:prediction-freightage}
\centering
 \footnotesize
\begin{tabular}{ccccccccccccccccc}
\hline
 Year    &Data         &{GM}(1,1)   &DGM(1,1)   &{${\rm GM_{SC}}$(1,1)} &{${\rm GM_{SD}}$(1,1) }  \\
 \hline
 2004    &5786         &5786.0000   &5786.0000  &5786.0000              &5786.0000         \\
 2005    &5973         &6015.9317   &6017.4333  &5973.0000              &5973.0000         \\
 2006    &6262         &6349.6357   &6351.3039  &6346.0123              &6361.0051    \\
 2007    &6840         &6701.8503   &6703.6990  &6724.0011              &6708.2685    \\
 2008    &7207         &7073.6022   &7075.6463  &7124.5042              &7138.8492    \\
 2009    &7332         &7465.9753   &7468.2307  &7548.8625              &7533.6395    \\ [4pt]
 2010    &8032         &7880.1133   &7882.5972  &7998.4969              &8012.2088    \\
 2011    &8882         &8317.2236   &8319.9544  &8474.9130              &8460.1700    \\
 2012    &9728         &8778.5804   &8781.5778  &8979.7061              &8992.7976    \\
 2013    &10531        &9265.5286   &9268.8140  &9514.5662              &9500.2731    \\
 2014    &11147        &9779.4880   &9783.0839  &10081.2842             &10093.7664   \\
\hline
\end{tabular}
\end{table}
\begin{table}[!htbp]
 \caption{Relative error values of freightage of Lanzhou by grey models (\%)}
 \label{table:err-freightage}
\centering
 \footnotesize
\begin{tabular}{ccccccccccccccccc}
\hline
 year    &{GM}(1,1)   &DGM(1,1)  &{${\rm GM_{SC}}$(1,1)}  &{${\rm GM_{SD}}$(1,1) }  \\
 \hline
 2004    &0           &0         &0                       &0        \\
 2005    &0.7188      &0.7439    &0                       &0        \\
 2006    &1.3995      &1.4261    &1.3416                  &1.5810   \\
 2007    &2.0197      &1.9927    &1.6959                  &1.9259   \\
 2008    &1.8509      &1.8226    &1.1447                  &0.9456   \\
 2009    &1.8273      &1.8580    &2.9578                  &2.7501   \\ [3pt]
 2010    &1.8910      &1.8601    &0.4171                  &0.2464   \\
 2011    &6.3587      &6.3279    &4.5833                  &4.7493   \\
 2012    &9.7597      &9.7288    &7.6922                  &7.5576   \\
 2013    &12.0166     &11.9854   &9.6518                  &9.7875   \\
 2014    &12.2680     &12.2357   &9.5606                  &9.4486   \\ [3pt]
 {${\rm MAPE_{\rm simu}}$}     &{\underline{\bf 1.5632}}      &1.5687    &1.8499     &1.8007   \\
 {${\rm MAPE_{\rm pred}}$}     &8.4588      &8.4276    &6.3810                  &{\underline{\bf 6.3579}} \\
 {${\rm MAPE_{\rm over}}$}     &5.0110      &4.9981    &4.3383                  &{\underline{\bf 4.3325}} \\
\hline
\end{tabular}
\end{table}

%
\section{Conclusions}
\label{sec:conclusion}

{The current study studied} the discrete GM(1,1) model with Simpson
formula called ${\rm GM_{SD}}$(1,1) model. Mathematical analysis is
presented to indicate the difference between the ${\rm
GM_{SD}}$(1,1) model and the ${\rm GM_{SC}}$(1,1) model. We also
proved our model is unbiased to simulate the homogeneous exponent
sequence. Applications are carried out to verify the performance of
our model with the other three models. {Computation results indicate
that ${\rm GM_{SD}}$(1,1) model provides accurate prediction,
outperforming GM(1,1), DGM(1,1) and ${\rm GM_{SC}}$(1,1) model. }

It may be remarked here that the idea for ${\rm GM_{SD}}$(1,1) model
used in our paper can be used to analyze other grey forecasting
model such as GM(1,$n$) or GMC(1,$n$) model. These are possible
extensions and suggested directions for our future research.

\section*{Acknowledgments}
\label{sec:acknowledgments}

This research was supported by the National Natural Science
Foundation of China (No.71901184, 71771033, 71571157, 11601357), the
Humanities and Social Science Project of Ministry of Education of
China (No.19YJCZH119), the Longshan academic talent research
supporting program of SWUST (No.17LZXY20), the Open Fund (PLN
201710) of State Key Laboratory of Oil and Gas Reservoir Geology and
Exploitation (Southwest Petroleum University), Applied Basic
Research Program of Science and Technology Commission Foundation of
Sichuan province (2017JY0159), the funding of V.C. \& V.R. Key Lab
of Sichuan Province (SCVCVR2018.08VS, SCVCVR2018.10VS), the Doctoral
Research Foundation of Southwest University of Science and
Technology (No.15zx7141, 16zx7140), and the National Statistical
Scientific Research Project (No.2018LY42).

\section*{References}

\bibliography{greybib}

\end{document}